\documentclass[12pt]{article}
\usepackage{amssymb,amsmath,epsfig}
\allowdisplaybreaks
\begin{document}
\title{\bf Ghost Dark Energy Model in $f(G)$ Gravity}
\author{M. Sharif \thanks{msharif.math@pu.edu.pk} and Saadia Saba
\thanks{saadia.saba86@gmail.com}\\
Department of Mathematics, University of the Punjab,\\
Quaid-e-Azam Campus, Lahore-54590, Pakistan.}
\date{}

\maketitle

\begin{abstract}
In this paper, we investigate cosmological consequences of ghost
dark energy model in modified Gauss-Bonnet gravity. We construct
ghost dark energy $f(G)$ model by using correspondence scenario for
both interacting and non-interacting schemes. For this purpose, we
consider FRW universe with pressureless matter and power-law scale
factor. We examine the behavior of equation of state parameter and
check the stability of ghost dark energy model through squared speed
of sound parameter. We also analyze the behavior of phase planes
like $\omega_{eff}-\omega'_{eff}$ and $r-s$ graphically. It is found
that the equation of state parameter represents quintessence era for
non-interacting while phantom phase of the universe for interacting
case. The squared speed of sound indicates stable ghost dark energy
model for both cases. The $\omega_{eff}-\omega'_{eff}$ plane shows
thawing region for non-interacting while freezing region for
interacting case. The $r-s$ plane corresponds to Chaplygin gas model
in both scenarios. We conclude that ghost dark energy model
describes evolution of the universe for appropriate choice of
parameters.
\end{abstract}

\section{Introduction}

The well-known phenomenon of accelerated expansion of the universe
is usually explained by the exotic type force known as dark energy
(DE). The cosmological constant is the simplest DE model and is the
base of $\Lambda$CDM model. Despite showing the consistent behavior
with all observational data, $\Lambda$CDM model undergoes several
difficulties like fine tuning and cosmic coincidence problem
\cite{2}. In order to substantiate the behavior of DE, researchers
used two different approaches, firstly the dynamical DE models
\cite{3} and secondly the modified gravity theories \cite{4}.

A dynamical DE model, known as Veneziano ghost DE (GDE) has been
introduced in late $70$'s by Veneziano \cite{5}. This has
significant non-trivial physical properties for the expanding
universe or in spacetime having non-trivial topological formation.
The existence of Veneziano GDE is supposed to resolve $U(1)$ problem
\cite{6}. The GDE has little contribution to the vacuum energy
density in a curved spacetime. It is proportional to
$\Lambda^{3}_{QCD}H$, where $\Lambda_{QCD}$ and $H$ are the quantum
chromodynamics (QCD) mass scale and Hubble parameter, respectively
\cite{7}. For measures $\Lambda_{QCD}\sim 100MeV$ and $H\sim
10^{-33}MeV$, $\Lambda^{3}_{QCD}H$ gives approximately $(10^{-3}
eV)^{4}$ to the observed DE density. This numeric value is
incredible to offer the necessary exotic force for accelerating
universe and also alleviates the fine tuning problem.

The Gauss-Bonnet (GB) theory is incredibly motivating theory as it
shows consistent behavior with solar system constraints. The GB
invariant has the expression as $G=R_{\alpha\rho\delta\sigma}
R^{\alpha\rho\delta\sigma}-4R_{\alpha\rho}R^{\alpha\rho}+R^{2}$,
where $R$, $R_{\alpha\rho}$ and $R_{\alpha\rho\delta\sigma}$ stand
for the Ricci scalar, Ricci tensor and Riemann curvature tensor,
respectively. This invariant is a four-dimensional topological
expression with restriction for spin-2 ghost instability. Nojiri and
Odintsov \cite{9} proposed $f(G)$ gravity by adding a generic
function in the Einstein-Hilbert action. This amazing theory
effectively describes the early and late-time cosmic evolution.
Cognola et al. \cite{10} studied DE model in modified GB gravity to
discuss cosmological evolution and also addressed the issues of
hierarchy problem. Nojiri et al. \cite{11} discussed DE in two
scenarios, one for implicit equation of state (EoS) of the universe
and other for modified GB gravity. They predicted the natural
transition from early to late-time universe. De Felice and Tsujikawa
\cite{12} analyzed consistency of the $f(G)$ model with solar system
constraints.

The current accelerated expansion of the universe can be
investigated by many DE models. Setare and Saridakis \cite{13}
explored the condition under which the holographic and GB DE models
describe the accelerated expansion of the universe. They also
studied the correspondence of holographic DE model with quintom,
phantom and canonical models and highlighted stable results for the
accelerated universe \cite{14}. Later, Setare et al. \cite{15}
implemented this concept of correspondence to different DE models
and modified theories. Sheykhi and Movahed \cite{16} discussed
implications of the interacting GDE model in general relativity and
observed expansion of the universe using constraints on the model
parameter. Sadeghi et al. \cite{16a} explored the interacting GDE
models by varying $G$ as well as $\Lambda$. They computed EoS and
deceleration parameters numerically to analyze the behavior of the
universe.

The reconstruction phenomenon in modified theories of gravity is a
useful technique to develop a viable DE model that anticipates the
history of cosmic evolution. This reconstruction scenario compares
the corresponding energy densities of DE model and modified theory
of gravity. In this scheme, we derive the modified generic function
of underlying theory through correspondence technique of energy
densities. Much work have been carried out in cosmology using this
scenario of correspondence for different DE models.

Saaidi et al. \cite{18} reconstructed the GDE $f(R)$ model using
correspondence scheme and analyzed its stability as well as
evolution by evaluating cosmological parameters. Alavirad and
Sheykhi \cite{17} studied cosmological constraints on GDE for FRW
universe using interaction between DE and dark matter in Brans-Dicke
theory. Fayaz and his collaborators \cite{19} investigated this
model by considering Bianchi-I universe in $f(R)$ gravity and
analyzed the cosmic evolution of the corresponding model.
Chattopadhyay \cite{20} explored stability of cosmic evolution using
cosmological parameters. Fayaz et al. \cite{23} investigated this
model in $f(R,T)$ gravity for Bianchi-I universe and concluded that
their results favor the current behavior of the universe.

The $f(G)$ gravity is an interesting modified gravity theory which
helps to better understand current and late-time acceleration of the
universe. Zhou et al. \cite{24} analyzed cosmological constraints of
DE model based on the modified GB gravity and derived the condition
of viability for the model with cosmic trajectories that mimics the
$\Lambda$CDM limit for both radiation as well as matter dominant
eras. Sheykhi and Bagheri \cite{25} explored quintessence GDE model
to describe recent evolution of the cosmos. Chattopadhyay \cite{26}
analyzed the generalized second law of thermodynamics in QCD ghost
$f(G)$ gravity. Shamir \cite{27} discussed viable DE models in
$f(G)$ gravity showing consistent behavior for the expansion of the
universe.

In this paper, we use correspondence scenario to reconstruct GDE
$f(G)$ model and examine the EoS parameter, squared speed of sound
parameter and phase planes. The format of the paper is as follows.
In the next section, we adopt reconstruction procedure for GDE
$f(G)$ model. Section \textbf{3} investigates evolution of the
universe for non-interacting case while section \textbf{4} examines
the interacting GDE $f(G)$ model. Finally, we discuss our results in
the last section.

\section{Reconstruction of GDE $f(G)$ Model}

In this section, we apply the correspondence between GDE and $f(G)$
gravity to reconstruct GDE $f(G)$ model. The action of $f(G)$
gravity is defined as \cite{28}
\begin{equation}\label{1}
S= \int d^{4}x \sqrt{-g}\left(\frac{R}{2\kappa^{2}}+f(G)+
\mathcal{L}_{m}\right),
\end{equation}
where $\kappa^{2}=1$ and $ \mathcal{L}_{m}$ are the coupling
constant and matter Lagrangian density, respectively. The
corresponding field equations are
\begin{equation}\label{1a}
R_{\alpha\beta}-\frac{1}{2} R g_{\alpha\beta}=  T_{\alpha\beta}^{eff},
\end{equation}
where $T_{\alpha\beta}^{eff}$ is the effective energy-momentum tensor
given by
\begin{eqnarray}\nonumber
T_{\alpha\beta}^{eff}
&=&\kappa^{2}T^{(m)}_{\alpha\beta}-8[R_{\alpha\rho\beta\gamma}+R_{\rho\beta}g_{\gamma\alpha}
+R_{\alpha\gamma}g_{\beta\rho}-R_{\gamma\rho}g_{\beta\alpha}
-R_{\alpha\beta}g_{\rho\gamma}\\\label{1b}&+&\frac{1}{2}R(g_{\alpha\beta}g_{\rho\gamma}-
g_{\rho\gamma}g_{\alpha\beta})]\nabla^{\rho}\nabla^{\gamma}f_{G}-
(Gf_{G}-f)g_{\alpha\beta},
\end{eqnarray}
where $f_{G}=\frac{df}{dG}$. Also, $\nabla_{\alpha}$ and
$T^{(m)}_{\alpha\beta}$ represent the covariant derivative and
matter energy-momentum tensor, respectively. The field equations for
FRW universe model in the presence of perfect fluid take the form
\begin{eqnarray}\label{1c}
3H^{2}=\rho_{m}+\rho_{DE},\quad-(2\dot{H}+3H^{2})=P_{m}+P_{DE},
\end{eqnarray}
where dot represents the time derivative and subscript $m$ denotes
matter contribution of energy density as well as pressure. The
energy density and pressure of dark source terms are
\begin{eqnarray}\label{2}
\rho_{DE}&=&\frac{1}{2}(Gf_{G}-f-24^{2}H^{4}(2\dot{H}^{2}+H\ddot{H}+4H^{2}
\dot{H})f_{GG}),\\\label{3}
P_{DE}&=&\frac{1}{2}(8H^{2}\ddot{f_{G}}+16H(H^{2}+\dot{H})\dot{f}_{G}-Gf_{G}+f),
\end{eqnarray}
where $G=24H^{2}(H^{2}+\dot{H})$.

The first field equation leads to
\begin{equation}\label{7}
\Omega_{m}+\Omega_{DE}=1,
\end{equation}
where $\Omega_{m}=\frac{\rho_{m}}{3H^{2}}$ and
$\Omega_{DE}=\frac{\rho_{DE}}{3H^{2}}$ are the fractional energy
densities associated with matter and dark source, respectively.
Dynamical DE models whose energy density is proportional to Hubble
parameter play a vital role in explaining accelerated expansion of
the universe. The GDE model is one of the dynamical DE model whose
energy density is defined as \cite{29}
\begin{equation}\label{8}
\rho_{GDE}=\alpha H,
\end{equation}
where $\alpha$ is an arbitrary constant having dimension
$[energy]^{3}$. We establish the correspondence between GDE and
$f(G)$ model by equating corresponding densities. Using
Eqs.(\ref{2}) and (\ref{8}), it follows that
\begin{equation}\label{9}
Gf_{G}-f-24^{2}H^{4}(2\dot{H}^{2}+H\ddot{H}+4H^{2}\dot{H})f_{GG}=2\alpha
H.
\end{equation}

In order to obtain the analytic solution of this equation, we
consider the following form of scale factor as
\begin{equation}\label{10}
a(t)=a_{0}t^{m},
\end{equation}
where $a_{0}$ is a constant representing the present day value of
the scale factor. Using Eq.(\ref{10}) in (\ref{9}), we obtain
\begin{equation}\label{11}
G^{2}f_{GG}+\frac{m-1}{4}Gf_{G}-\frac{m-1}{4}f= \frac{\alpha
m^{\frac{1}{4}}(m-1)^{\frac{3}{4}}G^{\frac{1}{4}}}{2^{\frac{7}{4}}
3^{\frac{1}{4}}},
\end{equation}
which is a second order linear differential equation whose solution
is
\begin{equation}\label{12}
f(G)=c_{1}G^{\frac{1}{4}(1-m)}+c_{2}G- \frac{\alpha
(m-1)^{\frac{3}{4}}G^{\frac{1}{4}}2^{\frac{9}{4}}}
{3^{\frac{5}{4}}m^{\frac{3}{4}}},
\end{equation}
where $c_{1}$ and $c_{2}$ are integration constants. This represents
the reconstructed GDE $f(G)$ model. Using Eq.(\ref{12}) in (\ref{2})
and (\ref{3}), we have
\begin{eqnarray}\label{16}
\rho_{DE}&=&\frac{\alpha m^{\frac{1}{4}}G^{\frac{1}{4}}}
{2^{\frac{3}{4}}3^{\frac{1}{4}}(m-1)^{\frac{1}{4}}},\\\label{17}
P_{DE}&=&\frac{\alpha G^{\frac{1}{4}}(1-3m)}
{2^{\frac{3}{4}}3^{\frac{5}{4}}m^{\frac{3}{4}}(m-1)^{\frac{1}{4}}},
\end{eqnarray}
where $m\neq1$. The graphical analysis of reconstructed GDE $f(G)$
model against $G$ is shown in Figure \textbf{1}. We take
$c_{1}=8.5$, $c_{2}=8.5$, $\alpha= 22.05$, $\rho_{m_{0}}=0.23$,
$a_{0}=1$, $\Omega_{m_{0}}=0.313$ and $H_{0}=67.48$ throughout the
analysis. It is observed that the reconstructed $f(G)$ model
initially exhibits rapidly decreasing behavior and then gradually
increases as $G$ increases in the range $1\leq m\leq3$.
\begin{figure}\center
\epsfig{file=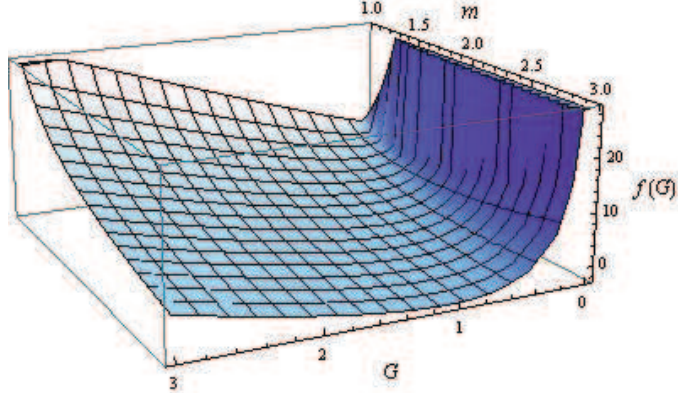,width=0.67\linewidth}\\
\caption{Plot of $f(G)$ for $\alpha=22.05$, $\Omega_{m_{0}}=0.313$
and $H_{0}=67.48$.}
\end{figure}

\section{Non-Interacting GDE $f(G)$ Model}

Here, we study non-interacting scenario of cold dark matter and GDE.
The conservation equations corresponding to matter and dark source
terms for pressureless fluid $(P_{m}=0)$ are
\begin{eqnarray}\label{4}
&&\dot{\rho}_{m}+3H\rho_{m}=0,\\\label{4a}
&&\dot{\rho}_{DE}+3H\rho_{DE}(1+\omega_{DE})=0.
\end{eqnarray}
Equation (\ref{4}) has solution of the form
\begin{equation}\label{5}
\rho_{m}=\rho_{m_{0}}a^{-3},
\end{equation}
where $\rho_{m_{0}}$ is an arbitrary constant.

In the following, we investigate the evolution of EoS parameter,
squared speed of sound and cosmological planes.

\subsection{The EoS Parameter}
\begin{figure}\center
\epsfig{file=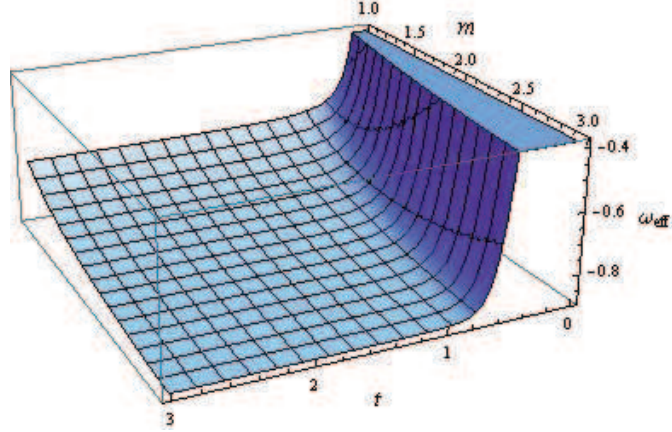,width=0.67\linewidth}\\
\caption{Plot of EoS parameter for $\alpha=22.05$,
$\rho_{m_{0}}=0.23$, $a_{0}=1$, $\Omega_{m_{0}}=0.313$ and
$H_{0}=67.48$.}
\end{figure}
The EoS parameter is given by
\begin{equation}\label{18}
\omega_{eff}=\frac{P_{eff}}{\rho_{eff}}=\frac{P_{DE}}{\rho_{DE}+\rho_{m}}.
\end{equation}
Using Eqs.(\ref{16}), (\ref{17}) and (\ref{5}) in the above
equation, we have
\begin{eqnarray}\label{19}
\omega_{eff}&=&\frac{-3m+1}{3m+\frac{3\rho_{m_{0}}t^{1-3m}}{\alpha
a_{0}^{3}}}.
\end{eqnarray}
Figure \textbf{2} shows graphical behavior of this parameter for $1<
m\leq3$. We observe that the EoS parameter presented quintessence
era and approaches the phantom divide line but never crosses it for
$m>1$ as $t$ increases. As the universe will collapse in the absence
of DE, thus the existence of DE elaborates our current accelerating
expansion of the universe. This shows that the GDE $f(G)$ model
favors DE phenomenon.

\subsection{Squared Speed of Sound Parameter}
\begin{figure}\center
\epsfig{file=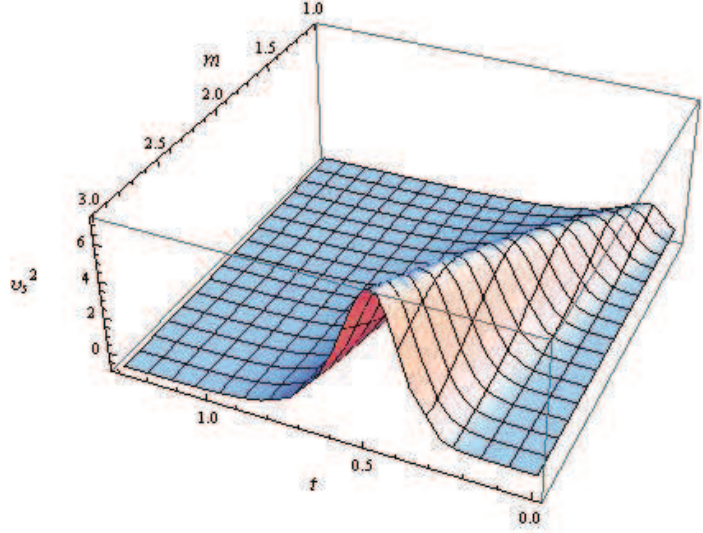,width=0.67\linewidth}\\
\caption{Plot of squared speed of sound parameter for
$\alpha=22.05$, $\rho_{m_{0}}=0.23$, $a_{0}=1$,
$\Omega_{m_{0}}=0.313$ and $H_{0}=67.48$.}
\end{figure}

We compute this parameter to analyze the stability of GDE $f(G)$
model. It has the following expression
\begin{equation}\label{20}
\nu_{s}^{2}= \frac{\dot{P}_{eff}}{\dot{\rho}_{eff}}.
\end{equation}
The sign of $\nu_{s}^{2}$ helps to determine stability of the
reconstructed DE model. A positive signature of $\nu_{s}^{2}$
designates stability of the model whereas its negative value
highlights the instability. Using Eqs.(\ref{16}) and (\ref{17}) in
(\ref{20}), it follows that
\begin{eqnarray}\nonumber
\nu_{s}^{2}&=&\frac{(1-3m)\alpha a_{0}^{3} m}{3m^{2}\alpha
a_{0}^{3}+\rho_{m_{0}}t^{1-3m}}(1+\frac{(1-3m)\rho_{m_{0}}t^{1-3m}}{3m^{2}\alpha
a_{0}^{3}+\rho_{m_{0}}t^{1-3m}})
\end{eqnarray}
We plot the squared speed of sound for $m>1$ as shown in Figure
\textbf{3}. It is observed that $\nu_{s}^{2}>0$ throughout the
evolution leading to the stable GDE $f(G)$ model.

\subsection{The $\omega_{eff}-\omega'_{eff}$ Plane}

Caldwell and Linder \cite{30} proposed $\omega_{eff}-\omega'_{eff}$
plane to analyze the behavior of quintessence DE model. They
classified the plane into two regions named as thawing
($\omega_{eff}<0,~ \omega'_{eff}>0$ ) and freezing regions
($\omega_{eff}<0,~ \omega'_{eff}<0$). Using Eq.(\ref{19}), we have
\begin{eqnarray}\nonumber
\omega'_{eff}=\frac{(1-3m)^{2}\rho_{m_{0}}t^{1-3m}\alpha
a_{0}^{3}}{(3m^{2}\alpha a_{0}^{3}+\rho_{m_{0}}t^{1-3m})^{2}}.
\end{eqnarray}
Figure \textbf{4} exhibits the $\omega_{eff}-\omega'_{eff}$ plane
for GDE $f(G)$ model with three distinct values of $m$, i.e.,
$m=2,2.4$ and $2.8$. It is found that $\omega_{eff}-\omega'_{eff}$
plane corresponds to thawing region for all considered values of $m$
showing a consistent behavior with our current accelerated expanding
universe.
\begin{figure}\center
\epsfig{file=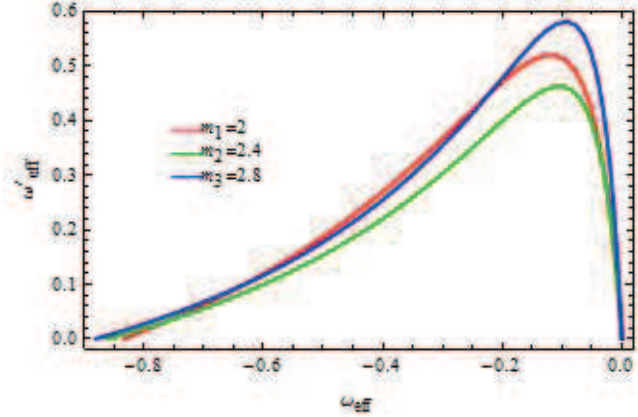,width=0.67\linewidth}\\
\caption{Trajectories of $\omega_{eff}-\omega'_{eff}$ for GDE $f(G)$
model with GDE parameter $\alpha=22.05$, $\rho_{m_{0}}=0.23$,
$a_{0}=1$, $\Omega_{m_{0}}=0.313$ and $H_{0}=67.48$.}
\end{figure}

\subsection{The $r-s$ Plane}

The accelerated expansion of the universe has been supported by
various DE models that represent the same values for deceleration
and Hubble parameters. These parameters fail to highlight the best
among those models. In this regard, Sahni et al. \cite{31}
introduced two dimensionless parameters in terms of deceleration and
Hubble parameters to classify DE models. These parameters are known
as statefinder parameters defined as
\begin{equation}\label{23}
r=\frac{\dddot{a}}{a H^{3}},\quad s=\frac{r-1}{3(q-\frac{1}{2})}.
\end{equation}
The parameter $r$ can be expressed in terms of deceleration
parameter as
\begin{equation}\label{24}
r=2 q^{2}+q-\acute{q}.
\end{equation}
These parameters help to determine the distance of a certain DE
model by using $\Lambda$CDM limit and also extricate the DE models.
They classified the universe in different regions, e.g., CDM limit
for $(r,s)=(1,0)$ and $\Lambda$CDM for $(r,s)=(1,1)$. Furthermore,
the region ($r<1,s>0$) interprets the phantom and quintessence DE
eras while the region $(s<0$, $r>1$) represents Chaplygin gas model.
Using Eq.(\ref{19}) in (\ref{23}) and (\ref{24}), we have
\begin{eqnarray}\nonumber
r&=&\frac{1}{2(3\alpha
m^{2}a_{0}^{3}t^{3m}+\rho_{m_{0}}t)^{2}}(18t^{6m}\alpha^{2}m^{4}a_{0}^{6}-15t^{3m+1}\alpha
m^{2}a_{0}^{3}\rho_{m_{0}}-27t^{6m}\alpha^{2}\\\nonumber&\times&m^{3}a_{0}^{6}+2t^{2}\rho_{m_{0}}^{2}+9t^{3m+1}
m\alpha\rho_{m_{0}}a_{0}^{3}+9t^{6m}\alpha^{2}m^{2}a_{0}^{6}+27t^{3m}\alpha
m^{3}\rho_{m_{0}}a_{0}^{3}\\\nonumber&-&18t^{3m}\alpha
m^{2}\rho_{m_{0}}a_{0}^{3}+3t^{3m}\alpha
m\rho_{m_{0}}a_{0}^{3}),\\\nonumber s&=&\frac{3}{4(3\alpha
m^{2}a_{0}^{3}t^{3m}+\rho_{m_{0}}t)^{3}}[(-1+3m)\alpha^{2}m^{2}a_{0}^{6}(\rho_{m_{0}}t^{6m}(9mt-9m^{2}+6m\\\nonumber&-&3t-1)+3\alpha
t^{9m}m a_{0}^{3}(3m-1))].
\end{eqnarray}
The trajectories of $r-s$ plane for GDE $f(G)$ model with $m=2,2.4$
and $2.8$ are shown in Figure \textbf{5}. These plots show that the
$r-s$ plane leads to the Chaplygin gas model regimes for all three
values of $m$ and meet the CDM limit while $\Lambda$CDM limit cannot
be obtained for the reconstructed model.
\begin{figure}\center
\epsfig{file=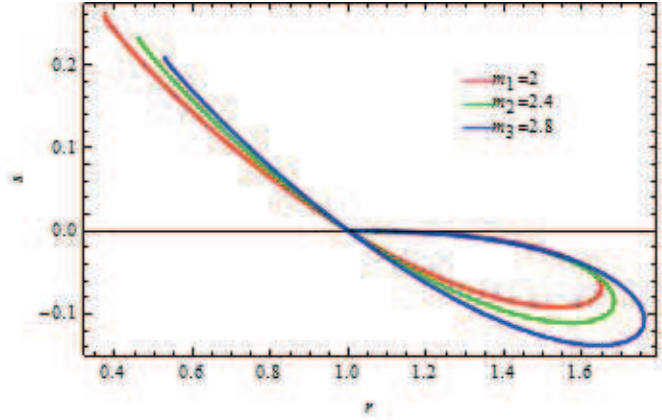,width=0.67\linewidth}\\
\caption{Trajectories of $r-s$ for GDE $f(G)$ model using
$\alpha=22.05$, $\rho_{m_{0}}=0.23$, $a_{0}=1$,
$\Omega_{m_{0}}=0.313$ and $H_{0}=67.48$.}
\end{figure}

\section{Interacting GDE $f(G)$ Model}

In this section, we investigate the interaction of GDE and
pressureless dark matter. Ghost DE and dark matter violate the
conservation equation while the interacting scenario leads to
\begin{eqnarray}\label{25}
&&\dot{\rho}_{m}+3H\rho_{m}=A,\\\label{26}
&&\dot{\rho}_{DE}+3H\rho_{DE}(1+\omega_{DE})=-A,
\end{eqnarray}
where $A$ is the interaction which transfers energies between CDM
and GDE. There are many simple choices like $3d_{1}H\rho_{DE}$,
$3d_{1}H\rho_{m}$, and $3d_{1}H(\rho_{DE}+\rho_{m})$ to describe
interaction in terms of energy densities and coupling constant
$d_{1}$. Cai and Su \cite{32} found that it is necessary for
interaction term to change its sign for evolving the universe from
deceleration to acceleration. However, the above three choices fail
to obey this condition. Therefore, we choose a particular form of
the interaction \cite{33} as
\begin{equation}\label{27}
A=3d_{1}H(\rho_{DE}-\rho_{m}),
\end{equation}
which changes its sign to describe the evolution of the universe
from deceleration to acceleration appropriately. Here, we analyze
some cosmological parameters for interacting GDE $f(G)$ model.

\begin{figure}\center
\epsfig{file=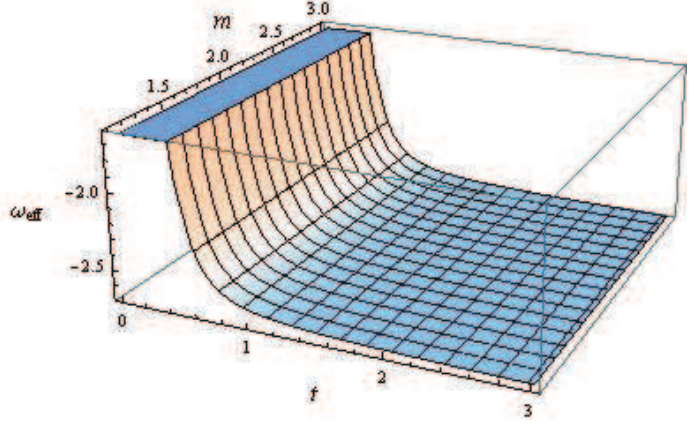,width=0.67\linewidth}\\
\caption{Plot of EoS parameter for $\alpha=22.05$,
$\rho_{m_{0}}=0.23$, $a_{0}=1$, $d_{1}=0.3$, $\Omega_{m_{0}}=0.313$
and $H_{0}=67.48$.}
\end{figure}
The EoS parameter is found by substituting Eqs.(\ref{1c}) and
(\ref{27}) in (\ref{25}) as follows
\begin{eqnarray}\nonumber
\omega_{eff}&=&(\frac{-3m+1}{3m}-d_{1}(\frac{-2}{\alpha
t}+\frac{-3m+1}{3m}+2))(1+\frac{\rho_{m_{0}}t^{1-3m+3d_{1}}}{\alpha
m a_{0}^{3}}\\\label{28}&+&\frac{3md_{1}}{1-3m+3md_{1}})^{-1}.
\end{eqnarray}
Figure \textbf{6} illustrates graphical description of EoS parameter
for $1< m<3$. It is observed that the EoS parameter shows phantom
phase of the universe for $m>1$ as $t$ increases.

The corresponding squared speed of sound parameter is given by
\begin{eqnarray}\nonumber
\nu_{s}^{2}&=&(\frac{-3m+1}{3m}-d_{1}(\frac{-2}{\alpha
t}+\frac{-3m+1}{3m}+2)+\frac{2d_{1}}{\alpha
t})(1+\frac{\rho_{m_{0}}t^{1-3m+3d_{1}}}{\alpha m
a_{0}^{3}}\\\nonumber&+&\frac{3md_{1}}{1-3m+3md_{1}})^{-1}+\frac{1}{\alpha
m a_{0}^{3}}[(\frac{-3m+1}{3m}-d_{1}(\frac{-2}{\alpha
t}+\frac{-3m+1}{3m}+2))\\\nonumber&\times&\rho_{m_{0}}t^{1-3m+3d_{1}}(1-3m+3d_{1})]
(1+\frac{\rho_{m_{0}}t^{1-3m+3d_{1}}}{\alpha m
a_{0}^{3}}+\frac{3md_{1}}{1-3m+3md_{1}})^{-2}.
\end{eqnarray}
We plot the squared speed of sound for the range $m>1$ as shown in
Figure \textbf{7}. It is observed that $\nu_{s}^{2}>0$ for coupling
constant $d_{1}=0.3$ leads to the stable GDE $f(G)$ model.
\begin{figure}\center
\epsfig{file=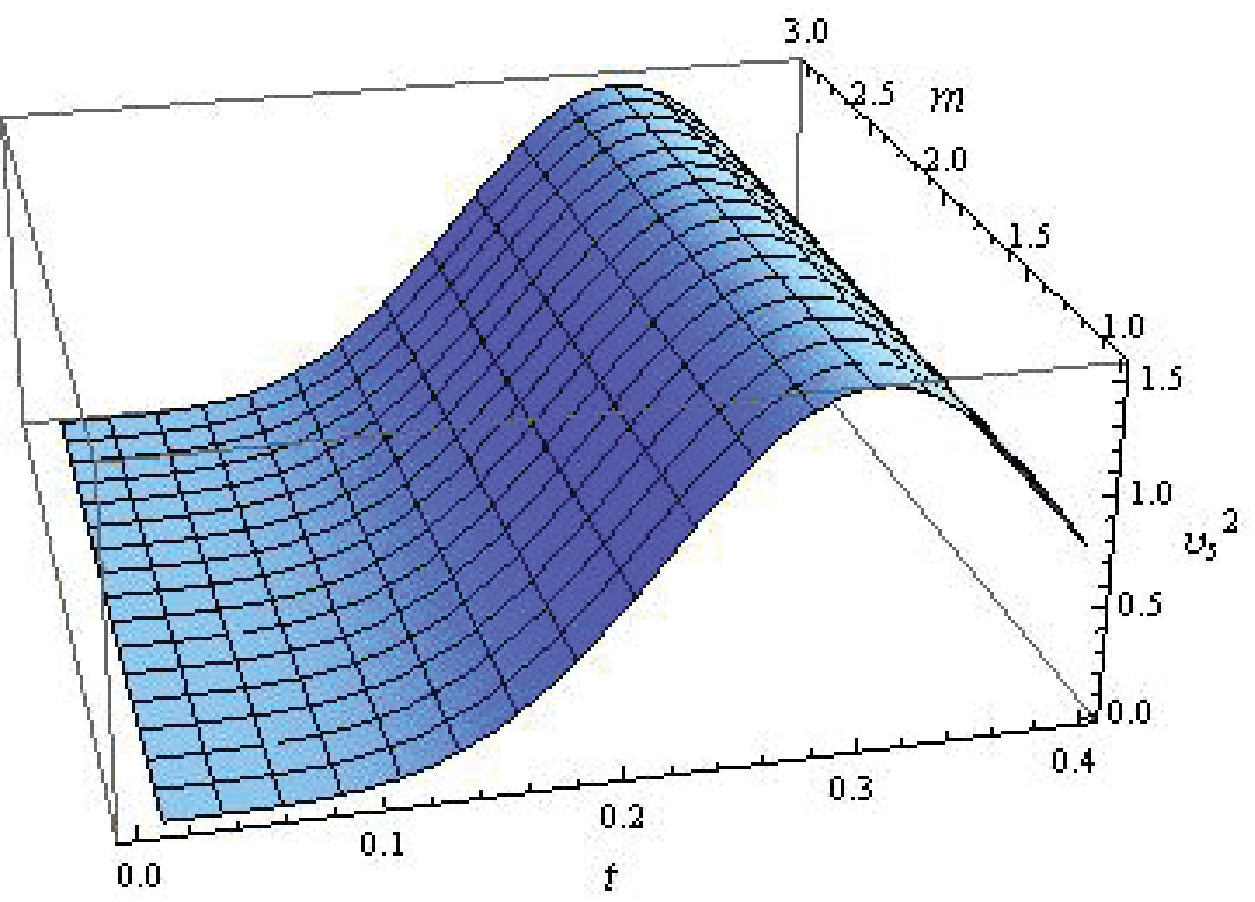,width=0.67\linewidth}\\
\caption{Plot of squared speed of sound parameter for
$\alpha=22.05$, $\rho_{m_{0}}=0.23$, $a_{0}=1$, $d_{1}=0.3$,
$\Omega_{m_{0}}=0.313$ and $H_{0}=67.48$.}
\end{figure}

The $\omega_{eff}-\omega'_{eff}$ plane can be found by using
Eq.(\ref{28}) as follows
\begin{eqnarray}\nonumber
\omega'_{eff}&=&\frac{-2d_{1}}{\alpha m
t}(1+\frac{\rho_{m_{0}}t^{1-3m+3d_{1}}}{\alpha m
a_{0}^{3}}+\frac{3md_{1}}{1-3m+3md_{1}})^{-1}-\frac{1}{\alpha m^{2}
a_{0}^{3}}[(\frac{-3m+1}{3m}-d_{1}\\\nonumber&\times&(\frac{-2}{\alpha
t}+\frac{-3m+1}{3m}+2))\rho_{m_{0}}t^{1-3m+3d_{1}}(1-3m+3d_{1})]
(1+\frac{\rho_{m_{0}}t^{1-3m+3d_{1}}}{\alpha m
a_{0}^{3}}\\\nonumber&+&\frac{3md_{1}}{1-3m+3md_{1}})^{-2}.
\end{eqnarray}
Figure \textbf{8} represents the $\omega_{eff}-\omega'_{eff}$ plane
for GDE $f(G)$ model with three distinct values of $m$, i.e.,
$m=2,2.4$ and $2.8$. It is found that $\omega_{eff}-\omega'_{eff}$
plane corresponds to freezing region for all considered values of
$m$ showing a consistent behavior with our current accelerated
expanding universe.
\begin{figure}\center
\epsfig{file=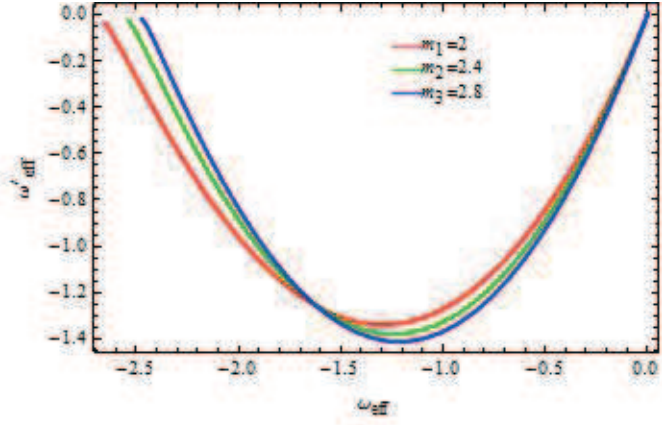,width=0.67\linewidth}\\
\caption{Trajectories of $\omega_{eff}-\omega'_{eff}$ for GDE $f(G)$
model with $\alpha=22.05$, $\rho_{m_{0}}=0.23$, $a_{0}=1$,
$d_{1}=0.3$, $\Omega_{m_{0}}=0.313$ and $H_{0}=67.48$.}
\end{figure}

\begin{figure}\center
\epsfig{file=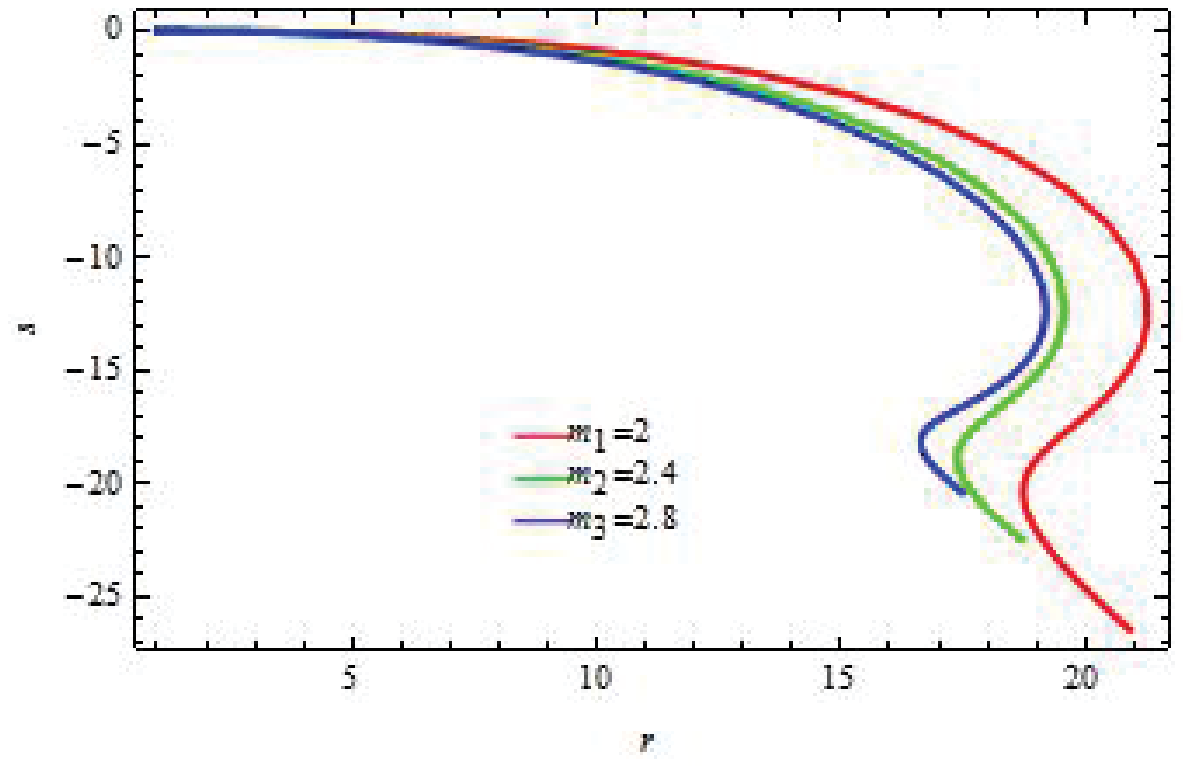,width=0.67\linewidth}\\
\caption{Trajectories of $r-s$ for GDE $f(G)$ model using
$\alpha=22.05$, $\rho_{m_{0}}=0.23$, $a_{0}=1$, $d_{1}=0.3$,
$\Omega_{m_{0}}=0.313$ and $H_{0}=67.48$.}
\end{figure}

The corresponding $r-s$ plane is given by
\begin{eqnarray}\nonumber
r&=&2[\frac{1}{2}+\frac{3}{2}((\frac{-3m+1}{3m}-d_{1}(\frac{-2}{\alpha
t}+\frac{-3m+1}{3m}+2))(1+\frac{\rho_{m_{0}}t^{1-3m+3d_{1}}}{\alpha
m a_{0}^{3}}\\\nonumber&+&\frac{3md_{1}}{1-3m+3md_{1}})^{-1})]^{2}+
\frac{1}{2}+\frac{3}{2}((\frac{-3m+1}{3m}-d_{1}(\frac{-2}{\alpha
t}+\frac{-3m+1}{3m}+2))\\\nonumber&\times&(1+\frac{\rho_{m_{0}}t^{1-3m+3d_{1}}}{\alpha
m
a_{0}^{3}}+\frac{3md_{1}}{1-3m+3md_{1}})^{-1})+\frac{3d_{1}}{\alpha
t^{2}}(1+\frac{\rho_{m_{0}}t^{1-3m+3d_{1}}}{\alpha m
a_{0}^{3}}\\\nonumber&+&\frac{3md_{1}}{1-3m+3md_{1}})^{-1}+\frac{3\rho_{m_{0}}
t^{1-3m+3d_{1}}(1-3m+3d_{1})}{2\alpha m a_{0}^{3} t
}((\frac{-3m+1}{3m}-d_{1}\\\nonumber&\times&(\frac{-2}{\alpha
t}+\frac{-3m+1}{3m}+2))(1+\frac{\rho_{m_{0}}t^{1-3m+3d_{1}}}{\alpha
m a_{0}^{3}}+\frac{3md_{1}}{1-3m+3md_{1}})^{-2}).\\\nonumber
s&=&\frac{1}{2}((\frac{-3m+1}{3m}-d_{1}(\frac{-2}{\alpha
t}+\frac{-3m+1}{3m}+2))(1+\frac{\rho_{m_{0}}t^{1-3m+3d_{1}}}{\alpha
m a_{0}^{3}}\\\nonumber&+&\frac{3md_{1}}{1-3m+3md_{1}})^{-1})
(2[\frac{1}{2}+\frac{3}{2}((\frac{-3m+1}{3m}-d_{1}(\frac{-2}{\alpha
t}+\frac{-3m+1}{3m}+2))\\\nonumber&\times&(1+\frac{\rho_{m_{0}}t^{1-3m+3d_{1}}}{\alpha
m a_{0}^{3}}+\frac{3md_{1}}{1-3m+3md_{1}})^{-1})]^{2}-
\frac{1}{2}+\frac{3}{2}((\frac{-3m+1}{3m}\\\nonumber&-&d_{1}(\frac{-2}{\alpha
t}+\frac{-3m+1}{3m}+2))(1+\frac{\rho_{m_{0}}t^{1-3m+3d_{1}}}{\alpha
m
a_{0}^{3}}+\frac{3md_{1}}{1-3m+3md_{1}})^{-1})\\\nonumber&+&\frac{3d_{1}}{\alpha
t^{2}}(1+\frac{\rho_{m_{0}}t^{1-3m+3d_{1}}}{\alpha m
a_{0}^{3}}+\frac{3md_{1}}{1-3m+3md_{1}})^{-1}+\frac{3\rho_{m_{0}}
t^{1-3m+3d_{1}}}{2\alpha m a_{0}^{3} t
}\\\nonumber&\times&(1-3m+3d_{1})((\frac{-3m+1}{3m}-d_{1}(\frac{-2}{\alpha
t}+\frac{-3m+1}{3m}+2))(1+\frac{\rho_{m_{0}}t^{1-3m+3d_{1}}}{\alpha
m a_{0}^{3}}\\\nonumber&+&\frac{3md_{1}}{1-3m+3md_{1}})^{-2})).
\end{eqnarray}
The trajectories of $r-s$ plane for GDE $f(G)$ model with $m=2,2.4$
and $2.8$ are shown in Figure \textbf{9}. These plots show that the
$r-s$ plane leads to the Chaplygin gas model regimes for all three
values of $m$.

\section{Concluding Remarks}

In this paper, we have used reconstruction scheme of GDE $f(G)$
model for both interacting as well as non-interacting scenario with
power-law form of the scale factor. We have established a
correspondence of GDE with $f(G)$ gravity and reconstructed $f(G)$
model by assuming GDE parameter $\alpha= 22.05$. To examine
cosmological behavior of the reconstructed $f(G)$ model, we have
discussed the cosmological parameters as well as phase planes. The
final results are summarized as follows.
\begin{itemize}
\item The reconstructed GDE $f(G)$ model (Figure \textbf{1}) represents
decreasing behavior initially and then it attains increasing
behavior forever. This shows that the reconstructed model is
a realistic one.
\item The EoS parameter (Figure \textbf{2}) shows
quintessence era of the universe for non-interacting case whereas it
represents phantom phase (Figure \textbf{6}) in interacting
scenario. Hence, our results are consistence with the current
accelerated cosmic behavior. We can conclude that the GDE $f(G)$
model favors the DE phenomenon.
\item The squared speed of sound parameter for both
interacting (Figures \textbf{3}) and non-interacting (\textbf{7})
cases indicates stability of the reconstructed model for current as
well as later epoch of time in the interval $1< m<3$ respectively.
\item The evolutionary behavior of the
$\omega_{DE}-\omega'_{DE}$ plane (Figures \textbf{4} and \textbf{8})
for $m=2,2.4$ and $2.8$ provides the thawing region for
non-interacting case whereas freezing region for interacting
scenario, respectively. Hence, cosmological expansion is more
accelerating in interacting case as compared with non-interacting
scenario.
\item The corresponding trajectories of $r-s$ plane indicate
Chaplygin gas model for all three values of $m$ in both cases.
Furthermore, it attains CDM limit but $\Lambda$CDM limit cannot be
achieved.
\end{itemize}

We have found that the GDE $f(G)$ model indicates stable behavior
and is consistent with the current behavior of the universe
depending on the appropriate choice of ghost parameter.
Chattopadhyay \cite{20} established the correspondence of $f(T)$
theory with GDE model and found that the EoS parameter never crosses
the phantom divide line in non-interacting scenario. Our results are
consistent with these outcomes. Saaidi et al. \cite{18} discussed
the correspondence between $f(R)$ theory and GDE model and found
that the reconstructed model is stable while the EoS parameter
passes through the phantom divide line for interacting case. Our
results are also consistent with these consequences. Finally, if we
take the coupling constant $d_{1}=0$ then all results of interaction
reduce to non-interacting scheme.

\end{document}